\documentclass[11pt]{article} 
\usepackage{amssymb} 
\usepackage{amsmath} 
\usepackage{amscd} 
\usepackage{latexsym} 
 
\catcode`@=11 
\renewcommand{\section}{\@startsection{section}{1}{0pt}{\medskipamount} 
{\medskipamount}{\large\bf}} 
\numberwithin{equation}{section} 
\catcode`@=12 
\def\a{\alpha} 
\def\b{\beta} 
\def\g{\gamma} 
\def\de{\delta} 
 
\def\ve{\varepsilon}
\def\z{\zeta} 
\def\h{\eta} 
\def\th{\theta}

\def\m{\mu} 
\def\n{\nu} 
\def\r{\rho} 
\def\s{\sigma}
\def\p{\phi} 
\def\vp{\varphi} 
\def\c{\chi} 
\def\j{\psi}

\def\La{\Lambda} 
\def\Ups{\Upsilon}

\def\da{\dot\alpha} 
\def\db{\dot\beta}
\def\dg{\dot\gamma}
\def\1{\dot 1} 
\def\2{\dot 2}

\newcommand{\C}{\mathbb C} 
\newcommand{\R}{\mathbb R}

\newcommand{\unity}{{\mathbf{1\hspace{-2.9pt}l}}}
\newcommand{\Hcal}{{\cal H}} 
\newcommand{\Acal}{{\cal A}}

\newcommand{\Ncal}{{\cal N}}
\newcommand{\Lcal}{{\cal L}} 
\newcommand{\Fcal}{{\cal F}}
 
\def\im{\textrm{i}} 
\def\e{\textrm{e}}
\def\N2{$N{=}2$} 
\def\pa{\partial}
\def\diff{\textrm{d}} 
\def\tr{\textrm{tr}} 
\def\sfrac#1#2{{\textstyle\frac{#1}{#2}}} 
\def\>{\rangle} 
\def\<{\langle} 
\def\+{\dagger} 
\def\={\ =\ } 
\def\und{\qquad\textrm{and}\qquad}

\begin{document} 
 
\begin{titlepage} 
\setcounter{page}{0} 
\begin{flushright} 
ITP--UH--09/07\\ 
\end{flushright} 
 
\vskip 2.0cm 
 
\begin{center} 
 
{\Large\bf Noncommutative Solitons in a \\[5mm]
           Supersymmetric Chiral Model in 2+1 Dimensions } 

\vspace{15mm}

{\large Olaf Lechtenfeld${}^1$} \ \ and \ \ {\large Alexander D. Popov${}^{1,2}$}
\\[5mm]
\noindent ${}^1${\em Institut f\"ur Theoretische Physik,
Leibniz Universit\"at Hannover \\
Appelstra\ss{}e 2, 30167 Hannover, Germany }
\\[5mm]
\noindent ${}^2${\em Bogoliubov Laboratory of Theoretical Physics, JINR\\
141980 Dubna, Moscow Region, Russia}
\\[5mm]
{Email: {\tt lechtenf, popov @itp.uni-hannover.de}}

\vspace{15mm}

\begin{abstract} 
\noindent 
We consider a supersymmetric Bogomolny-type model in 2+1 dimensions 
originating from twistor string theory. By a gauge fixing this model 
is reduced to a modified U($n$) chiral model with $2\Ncal{\le}\,8$ 
supersymmetries in 2+1 dimensions. After a Moyal-type deformation of the 
model, we employ the dressing method to explicitly construct multi-soliton 
configurations on noncommutative $\R^{2,1}$ and analyze some of their 
properties.
\end{abstract} 
 
\end{center} 
\end{titlepage}

\section{Introduction} 
 
\noindent
In the low-energy limit string theory with D-branes gives rise to
noncommutative field theory on the branes when the string propagates in a 
nontrivial NS-NS two-form ($B$-field) background~\cite{DH, Chu, Sch, SW}. 
In particular, if the open string has $N{=}2$ worldsheet supersymmetry, 
the tree-level target space dynamics is described by a noncommutative 
self-dual Yang-Mills (SDYM) theory in 2+2 dimensions~\cite{LPS1}. 
Furthermore, open $N{=}2$ strings in a $B$-field background induce on the 
worldvolume of $n$ coincident D2-branes a noncommutative Yang-Mills-Higgs 
Bogomolny-type system in 2+1 dimensions which is equivalent to a 
noncommutative generalization~\cite{LPS2} of the modified U($n$) chiral 
model known as the Ward model~\cite{Ward88}.
The topological nature of $N{=}2$ strings and the integrability of their 
tree-level dynamics~\cite{OV} render this noncommutative sigma model 
integrable.\footnote{
For discussing some other noncommutative integrable models 
see e.g.~\cite{group5, Haman} and references therein.}

Being integrable, the commutative U($n{\ge}2$) Ward model features a plethora 
of exact scattering and no-scattering multi-soliton and wave solutions, 
i.e.~time-dependent stable configurations on~$\R^2$. These are not only
a rich testing ground for physical properties such as adiabatic dynamics or
quantization, but also descend to more standard multi-solitons of various
integrable systems in 2+0 and 1+1 dimensions, such as sine-Gordon,
upon dimensional and algebraic reduction. There is a price to pay however:
Nonlinear sigma models in 2+1 dimensions
may be Lorentz-invariant or integrable but not both~\cite{Ward88, Zakr}.
In fact, Derrick's theorem prohibits the existence of stable solitons
in Lorentz-invariant scalar field theories above 1+1 dimensions.
A Moyal deformation, however, overcomes this hurdle, but of course
replaces Lorentz invariance by a Drinfeld-twisted version. 
There is another gain: The deformed Ward model possesses not only deformed
versions of the just-mentioned multi-solitons, but in addition allows for
a whole new class of genuinely noncommutative (multi-)solitons, 
in particular for the U(1)~group~\cite{LP1, LP2}! Moreover, 
this class is related to the generic but perturbatively constructed 
noncommutative scalar-field solitons~\cite{gomistr, goheaspr} 
by an infinite-stiffness limit of the potential~\cite{LKP}.

In~\cite{LP1, LP2} and \cite{B}--\cite{Penati} families of multi-solitons 
as well as their reduction to solitons of the noncommutative sine-Gordon 
equations were described and studied. In the nonabelian case both scattering 
and nonscattering configurations were obtained. For static configurations the 
issue of their stability was analyzed~\cite{DLP}. The full moduli space metric
for the abelian model was computed and its adiabatic two-soliton dynamics was 
discussed~\cite{LKP}.

Recall that the critical $N{=}2$ string theory has a four-dimensional target 
space, and its open string effective field theory is self-dual 
Yang-Mills~\cite{OV}, which gets deformed noncommutatively in the presence 
of a $B$-field~\cite{LPS1}. Conversely, the noncommutative SDYM equations are 
contained~\cite{IU} in the equations of motion of  $N{=}2$ string field theory
(SFT)~\cite{Berk} in a $B$-field background. This SFT formulation is based on 
the $N{=}4$ topological string description~\cite{BeVa}. It is well known that 
the SDYM model can be described in terms of holomorphic bundles over (an open 
subset of) the twistor space\footnote{
For reviews of twistor theory see, e.g., the books~\cite{WW, MW}.}
\cite{Ward77} $\C P^3$ and the topological $N{=}4$ string theory contains 
twistors from the outset. The Lax pair, integrability and the solutions to 
the equations of motion by twistor and dressing methods were incorporated 
into the $N{=}2$ open SFT in~\cite{group1, group2}. However, this theory 
reproduces only {\it bosonic\/} SDYM theory, its symmetries 
(see e.g.~\cite{group4, group4a, IvLe00}) and integrability properties. 
It is natural to ask: What string theory can describe 
{\it supersymmetric\/} SDYM theory~\cite{SemVol, Sieg} in four dimensions?

There are some proposals~\cite{Sieg, group8, BeSi, BeGaLe} for extending 
$N{=}2$ open string theory (and its SFT) to be space-time supersymmetric. 
Moreover, it was shown by Witten~\cite{Wit} that ${\Ncal}{=}4$ supersymmetric 
SDYM theory appears in twistor string theory, which is a B-type open 
topological string with the supertwistor space $\C P^{3|4}$ as a target 
space.\footnote{
For other variants of twistor string models see~\cite{group3,group3a,group3b}.
For recent reviews providing a twistor description of super Yang-Mills theory, 
see~\cite{PoSaWo, SaWo} and references therein.}
Note that ${\Ncal}{<}4$ SDYM theory forms a BPS subsector of 
${\Ncal}$-extended super Yang-Mills theory, and ${\Ncal}{=}4$ SDYM can be 
considered as a truncation of the full ${\Ncal}{=}4$ super Yang-Mills 
theory~\cite{Wit}. It is believed~\cite{NeVa, group3a} that twistor string 
theory is related with the previous proposals~\cite{Sieg,group8,BeSi,BeGaLe} 
for a Lorentz-invariant supersymmetric extension of $N{=}2$ (and topological 
$N{=}4$) string theory which also leads to the ${\Ncal}{=}4$ SDYM model. 

A dimensional reduction of the above relations between twistor strings and 
${\Ncal}{=}4$ super Yang-Mills and SDYM models was considered 
in~\cite{group9, PSW, Sa, sigma8}. The corresponding twistor string theory 
after this reduction is the topological B-model on the mini-supertwistor 
space~${\cal P}^{2|4}$.
In~\cite{sigma8} it was shown that the ${2\Ncal}{=}8$ supersymmetric 
extension of the Bogomolny-type model in 2+1 dimensions is equivalent to an 
${2\Ncal}{=}8$ supersymmetric modified U($n$) chiral model on $\R^{2,1}$.
The subject of the current paper is an ${2\Ncal}{\le}8$ version of
the above supersymmetric Bogomolny-type Yang-Mills-Higgs model in signature 
$(-++)$, its relation with an ${\Ncal}$-extended supersymmetric modified 
integrable U($n$) chiral model (to be defined) in 2+1 dimensions and
the Moyal-type noncommutative deformation of this chiral model. 
We go on to explicitly construct multi-soliton configurations on noncommutative 
$\R^{2,1}$ for the corresponding supersymmetric sigma model field equations. 
By studying the scattering properties of the constructed configurations, 
we prove their asymptotic factorization without scattering for large times. 
We also briefly discuss a D-brane interpretation of these soliton 
configurations from the viewpoint of twistor string theory.

\vspace{5mm} 
 
\section{Supersymmetric Bogomolny model in 2+1 dimensions} 
 
\noindent
{\bf 2.1 \ $\Ncal$-extended SDYM equations in 2+2 dimensions}

\smallskip

\noindent
{\bf Space $\R^{2,2}$. } Let us consider the four-dimensional space 
$\R^{2,2}=(\R^4, g)$ with the metric
\begin{equation}\label{s2} 
\diff s^2 \= g_{\mu\nu}\diff x^\mu\diff x^\nu \=\det (\diff x^{\a\da})\=
\diff x^{1\1}\diff x^{2\2} -
\diff x^{2\1}\diff x^{1\2}
\end{equation} 
with $(g_{\mu\nu})=\textrm{diag}(-1,+1,+1,-1)$, 
where $\mu,\nu,\ldots=1,\ldots,4$ are 
space-time indices and $\a = 1,2$, $\da = \1 , \2 $ are spinor indices. We 
choose the coordinates\footnote{Our conventions are chosen to match those 
of~\cite{LP1} after reduction to the space $\R^{2,1}$ with coordinates 
$(t,x,y)$.}
\begin{equation}\label{xmu} 
(x^\mu ) \= (x^a , \tilde t) \=(t,x,y,\tilde t)
\qquad\textrm{with}\qquad a,b,\ldots=1,2,3  \ ,
\end{equation}
and the signature $(-++\,-)$ allows us to introduce real isotropic coordinates 
(cf.~\cite{IU, LPS2})
\begin{equation}\label{iso} 
x^{1\1} = \sfrac12(t-y)\ ,\quad x^{1\2} = \sfrac12(x+\tilde t)\ ,\quad 
x^{2\1} = \sfrac12(x-\tilde t)\ ,\quad x^{2\2} = \sfrac12(t + y)  \ . 
\end{equation}

\smallskip

\noindent
{\bf SDYM. } Recall that the SDYM equations for a field strength 
tensor~$F_{\mu\nu}$ on $\R^{2,2}$ read
\begin{equation}\label{sdym} 
\sfrac12\ve_{\mu\nu\r\s} F^{\r\s}=F_{\m\n}\ , 
\end{equation}
where $\ve_{\mu\nu\r\s}$ is a completely antisymmetric tensor on $\R^{2,2}$ 
and $\ve_{1234}=1$. In the coordinates (\ref{iso}) we have the decomposition 
\begin{equation}\label{Fada} 
F_{\a\da ,\b\db} \= 
\pa _{\a\da}A_{\b\db} - \pa_{\b\db} A_{\a\da} + [A_{\a\da}, A_{\b\db}]
\= \ve_{\a\b}\,F_{\da\db} + \ve_{\da\db}\,F_{\a\b} 
\end{equation}
with
\begin{equation}\label{fab}
F_{\da\db}\ :=\ -\sfrac12\ve^{\a\b} F_{\a\da ,\b\db} \und
F_{\a\b}  \ :=\ -\sfrac12\ve^{\da\db} F_{\a\da ,\b\db}\ ,
\end{equation}
where $\ve_{\a\b}$ is antisymmetric, $\ve_{\a\b}\ve^{\b\g}=\de_\a^\g$, 
and similar for $\ve^{\da\db}$, with $\ve^{12}=\ve^{\1\2}=1$.
The gauge potential $(A_{\a\da})$ will appear in the covariant derivative
\begin{equation}
D_{\a\db} \= \pa_{\a\db} + [A_{\a\db},\ \cdot\ ] \ .
\end{equation}
In spinor notation, (\ref{sdym}) is equivalently written as
\begin{equation}\label{spinorsdym}
F_{\da\db}\=0 \qquad\Leftrightarrow\qquad
F_{\a\da ,\b\db} \= \ve_{\da\db}\,F_{\a\b} \ .
\end{equation} 
Solutions $\{A_{\a\da}\}$ to these equations form a subset (a BPS sector) of the
solution space of Yang-Mills theory on $\R^{2,2}$.

\smallskip

\noindent
{\bf $\Ncal$-extended SDYM in component fields. } The field content of 
$\Ncal$-extended super SDYM is\footnote{We use symmetrization $(\cdot )$ 
and antisymmetrization $[\cdot ]$ of $k$ indices
with weight $\frac{1}{k!}$, e.g. $[ij]=\frac{1}{2!}(ij-ji)$.}
\begin{subequations}\label{1-5}
\begin{eqnarray}
{\Ncal}=0 & & A_{\alpha\da}\\
{\Ncal}=1 & & A_{\alpha\da},\ \chi^i_{\a}
\qquad\textrm{with}\quad i=1\\
{\Ncal}=2 & & A_{\a\da},\ \chi^i_{\a},\ \phi^{[ij]}
\qquad\textrm{with}\quad i,j=1,2\\
{\Ncal}=3 & &A_{\a\da},\ \chi^i_{\a},\ \phi^{[ij]},\ \tilde\chi^{[ijk]}_{\da} 
\qquad\textrm{with}\quad i,j,k=1,2,3\\
\Ncal=4 & & A_{\a\da},\ \chi^i_{\a},\ \phi^{[ij]},\ \tilde\chi^{[ijk]}_{\da},\ 
      G_{\da\db}^{[ijkl]}
\qquad\textrm{with}\quad i,j,k,l=1,2,3,4 \ .
\end{eqnarray}
\end{subequations}
Here $(A_{\a\da},\ \chi^i_{\a},\ \phi^{[ij]},\ \tilde\chi^{[ijk]}_{\da} ,\ 
G_{\da\db}^{[ijkl]})$ are fields of helicities $(+1,+\sfrac12,0,-\sfrac12,-1)$.
These fields obey the field equations of the $\Ncal=4$ SDYM 
model, namely~\cite{Sieg, Wit}
\begin{subequations}\label{SDYMeom2}
\begin{eqnarray}
&&F_{\da\db}\=0\ ,\\[2pt]
&&D_{\a\da}\c^{i\a}\=0\ ,\\[2pt]
&&D_{\a\da}D^{\a\da}\p^{ij}+2\{\c^{i\a},\c^j_{\a}\}\=0\ ,\\[2pt]
&&D_{\a\da}\tilde{\c}^{\da [ijk]}- 6[\c^{[i}_{\a},\p^{jk]}]\=0\ ,\\[2pt]
&&D_{\a}^{\ \dg}G_{\dg\db}^{[ijkl]}+12\{\c^{[i}_\a,\tilde\c_{\db}^{jkl]}\}
  -18[\p^{[ij},D_{\a\db}\p^{kl]}]\=0\ .
\end{eqnarray}
\end{subequations}
Note that the $\Ncal<4$ SDYM field equations 
are governed by the first $\Ncal{+}1$ equations of (\ref{SDYMeom2}), where 
$F_{\da\db}=0$ is counted as one equation and so on.
 
\vspace{5mm} 
\newpage 
 
\noindent
{\bf 2.2 \ Superfield formulation of $\Ncal$-extended SDYM}

\smallskip

\noindent
{\bf Superspace $\R^{4|4\Ncal}$. } Recall that in the space $\R^{2,2}=(\R^4,g)$
with the metric $g$ given in (\ref{s2})  one may introduce {\it purely real\/}
Majorana-Weyl spinors\footnote{
Note that in Minkowski signature the Weyl spinor
$\th^\a$ is complex and $\h_{\da}=\ve_{\da\db}\h^{\db} =\overline{\th^\a}$ is
complex conjugate to~$\th^\a$. For the Kleinian (split) signature $2+2$, 
however, these spinors are real and independent of one another.} 
$\th^\a$ and $\h^{\da}$ of helicities $+\sfrac12$ and $-\sfrac12$ 
as anticommuting (Grassmann-algebra) objects.
Using $2\Ncal$ such spinors with components $\th^{i\a}$ and $\h_i^{\da}$ 
for $i=1,\ldots,\Ncal$, one can define the $\Ncal$-extended superspace 
$\R^{4|4\Ncal}$ and the $\Ncal$-extended supersymmetry algebra generated by the
supertranslation operators
\begin{equation}\label{2.12}
P_{\a\da}\=\pa_{\a\da}\ ,\qquad 
Q_{i\a}\=\pa_{i\a}-\h_i^{\da}\pa_{\a\da} \und
Q^{i}_{\da}\=\pa^i_{\da}-\th^{i\a}\pa_{\a\da}\ ,
\end{equation}
where
\begin{equation}
\pa_{\a\da}\ :=\ \frac{\pa}{\pa x^{\a\da}}\ ,\qquad 
\pa_{i\a}\ :=\ \frac{\pa}{\pa \th^{i\a}} \und
\pa^i_{\da}\ :=\ \frac{\pa}{\pa \h^{\da}_i}\ .
\end{equation}
The commutation relations for the generators (\ref{2.12}) read
\begin{equation}
\{Q_{i\a}, Q^{j}_{\da}\} \= -2\de_i^j P_{\a\da}\ , \qquad 
[P_{\a\da}, Q_{i\b}] \= 0 \und
[P_{\a\da}, Q^i_{\db}] \= 0\ .
\end{equation}

To rewrite equations of motion in terms of $\R^{4|4\Ncal}$ superfields 
one uses the additional operators
\begin{equation}
D_{i\a} \= \pa_{i\a}+\h_i^{\da}\pa_{\a\da} \und
D^{i}_{\da} \= \pa^i_{\da}+\th^{i\a}\pa_{\a\da}\ ,
\end{equation}
which (anti)commute with the operators (\ref{2.12}) and satisfy
\begin{equation}
\{D_{i\a}, D^{j}_{\db}\} \= 2\de_i^j P_{\a\db}\ ,\qquad 
[P_{\a\da}, D_{i\b}] \= 0 \und
[P_{\a\da}, D^j_{\db}] \= 0\ .
\end{equation}

\smallskip

\noindent
{\bf Antichiral superspace $\R^{4|2\Ncal}$. } On the superspace $\R^{4|4\Ncal}$ 
one may introduce tensor fields depending on bosonic and 
fermionic coordinates (superfields), differential forms, Lie 
derivatives ${\Lcal}_X$ etc.. Furthermore, on any such superfield $\Acal$ one 
can impose the constraint equations $\Lcal_{D_{i\a}}\Acal =0$, which for 
a scalar superfield $f$ reduce to the so-called antichirality conditions
\begin{equation}\label{Df}
{D_{i\a}}f\=0\ .
\end{equation}
These are easily solved by using a coordinate transformation 
on $\R^{4|4\Ncal}$,
\begin{equation}\label{xht}
(x^{\a\da},\ \h_i^{\da},\ \th^{i\a})\quad\to\quad(\tilde x^{\a\da}=x^{\a\da}{-}
\th^{i\a}\h_i^{\da},\ \h_i^{\da},\ \th^{i\a})\ ,
\end{equation}
under which $\pa_{\a\da}, D_{i\a}$ and $D^i_{\da}$ transform to the operators
\begin{equation}\label{pDD}
\tilde\pa_{\a\da}\=\pa_{\a\da}\ ,\qquad 
\tilde D_{i\a}\=\pa_{i\a} \und
\tilde D^i_{\da}\=\pa^i_{\da}+ 2\th^{i\a}\pa_{\a\da}\ .
\end{equation}
Then (\ref{Df}) simply means that $f$ is defined on a sub-superspace 
$\R^{4|2\Ncal}\subset\R^{4|4\Ncal}$ with coordinates
\begin{equation}\label{xh}
\tilde x^{\a\da} \und \h_i^{\da}\ .
\end{equation}
This space is called antichiral superspace. In the following we will usually 
omit the tildes when working on the antichiral superspace.

\smallskip

\noindent
{\bf $\Ncal$-extended SDYM in superfields. } The $\Ncal$-extended 
SDYM equations can be rewritten in terms of superfields on the antichiral 
superspace $\R^{4|2\Ncal}$~\cite{Sieg, DevOg}. 
Namely, for any given $0\le\Ncal\le 4$, 
fields of a proper multiplet from (\ref{1-5})
can be combined into superfields $\Acal_{\a\da}$ and $\Acal^i_{\da}$ 
depending on $x^{\a\da},\h_i^{\da}\in  \R^{4|2\Ncal}$ and giving rise to
covariant derivatives
\begin{equation}\label{covder}
\nabla_{\alpha\da}\ :=\ \pa_{\alpha\da} +\Acal_{\alpha\da} \und
\nabla_{\da}^i\ :=\ \pa_{\da}^i+\Acal_{\da}^i\ .
\end{equation}
In such terms the $\Ncal$-extended SDYM equations (\ref{SDYMeom2}) read
\begin{equation}\label{sdym1}
[\nabla_{\a\da},\nabla_{\b\db}]+[\nabla_{\a\db},\nabla_{\b\da}]=0\ ,\quad
[\nabla_{\da}^i,\nabla_{\b\db}]+[\nabla_{\db}^i,\nabla_{\b\da}]=0\ ,\quad
\{\nabla_{\da}^i,\nabla_{\db}^j\}+\{\nabla_{\db}^i, \nabla_{\da}^j\}=0\ ,
\end{equation}
which is equivalent to 
\begin{equation}\label{sdym2}
[\nabla_{\a\da},\nabla_{\b\db}]\=\ve_{\da\db}\,\Fcal_{\a\b}\ ,\qquad 
[\nabla^i_{\da},\nabla_{\b\db}]\=\ve_{\da\db}\,\Fcal^i_{\b} \und
\{\nabla^i_{\da},\nabla^j_{\db}\}\=\ve_{\da\db}\,\Fcal^{ij}\ ,
\end{equation}
where $\Fcal^{ij}$ is antisymmetric and $\Fcal_{\alpha\beta}$ is symmetric
in their indices. 

The above gauge potential superfields $(\Acal_{\a\da},\ \Acal^i_{\da})$ as well 
as the gauge strength superfields $(\Fcal_{\a\b},\ \Fcal^i_\a,\ \Fcal^{ij})$ 
contain all physical component fields of the $\Ncal$-extended SDYM model. 
For instance, the lowest component of the triple 
$(\Fcal_{\a\b},\ \Fcal^i_\a,\ \Fcal^{ij})$ in an $\h$-expansion is
$(F_{\a\b},\ \chi^i_\a,\ \p^{ij})$, with zeros in case $\Ncal$ is too small.
By employing Bianchi identities for the gauge strength superfields, 
one successively obtains~\cite{DevOg} the superfield expansions and the 
field equations (\ref{SDYMeom2}) for all component fields.

It is instructive to extend the antichiral combination in (\ref{pDD}) to 
potentials and covariant derivatives,
\begin{equation}\label{chiralcovariant}
\begin{matrix}
\tilde D^i_{\da} &=& \pa^i_{\da} &+& 2\,\th^{i\a}\,\pa_{\a\da} \\[4pt]
+ & & + & & \phantom{XX} + \\[4pt]
\tilde\Acal^i_{\da} &:=& \Acal^i_{\da} &+& 2\,\th^{i\a}\,\Acal_{\a\da} \\[4pt]
\parallel & & \parallel & & \phantom{XX} \parallel \\[4pt]
\tilde\nabla^i_{\da} &:=& \nabla^i_{\da} &+& 2\,\th^{i\a}\,\nabla_{\a\da}
\end{matrix} 
\end{equation}
where $\nabla_{\alpha\da}$, $\nabla^i_{\da}$ and $\tilde D^i_{\da}$ are given 
by (\ref{covder}) and (\ref{pDD}), while $\Acal^i_{\da}$ and $\Acal_{\a\da}$ 
depend on $x^{\a\da}$ and $\h_i^{\da}$ only.
With the antichiral covariant derivatives, 
one may condense (\ref{sdym1}) or (\ref{sdym2}) into the single set
\begin{equation}\label{sdym3}
\{\tilde\nabla^i_{\da}, \tilde\nabla^j_{\db}\}\ +\ 
\{\tilde\nabla^i_{\db}, \tilde\nabla^j_{\da}\}\=0
\qquad\Leftrightarrow\qquad
\{\tilde\nabla^i_{\da},\tilde\nabla^j_{\db}\}\=\ve_{\da\db}\,\tilde\Fcal^{ij}\ ,
\end{equation}
with $\tilde\Fcal^{ij}=\Fcal^{ij}+4\,\th^{[i\a}\Fcal^{j]}_{\a}+
4\,\th^{i\a}\th^{j\b}\Fcal_{\a\b}$.
The concise form (\ref{sdym3}) of the  
$\Ncal$-extended SDYM equations is quite convenient, and we will use it 
interchangeable with (\ref{sdym1}).

\smallskip

\noindent
{\bf Linear system for $\Ncal$-extended SDYM. }
It is well known that the superfield SDYM equations (\ref{sdym1}) can be seen
as the compatibility conditions for the linear system of differential equations
\begin{equation}\label{ls}
\zeta^{\da}(\pa_{\a\da}+\Acal_{\a\da})\,\psi \= 0 \und
\zeta^{\da}(\pa_{\da}^i+\Acal_{\da}^i)\,\psi \= 0\ ,
\end{equation}
where $(\z_{\db})=\bigl(\begin{smallmatrix}1\\ \z\end{smallmatrix}\bigr)$ 
and $\z^{\da}=\ve^{\da\db}\z_{\db}.$ 
The extra (spectral) parameter\footnote{The parameter $\zeta$ is related 
with $\lambda$ used in~\cite{PSW} by the formula $\zeta=\im\frac{1-
\lambda}{1+\lambda}$ (cf. e.g.~\cite{IvLe00}).} $\zeta$ lies in the extended 
complex plane $\C\cup\infty = \C P^1$. Here $\j$ is a matrix-valued function 
depending not only on $x^{\a\da}$  and  $\h_i^{\da}$ but also (meromorphically) 
on $\zeta\in\C P^1$. We subject the $n{\times}n$ matrix $\j$ to the 
following reality condition:
\begin{equation}\label{recon}
\j (x^{\a\da}, \h_i^{\da}, \zeta )\left [\j (x^{\a\da}, \h_i^{\da}, 
\bar\zeta)\right ]^\+ \=\unity\ ,
\end{equation} 
where ``$\+$" denotes hermitian conjugation and $\bar\zeta$ is complex 
conjugate to~$\zeta$. This condition guarantees that all physical fields 
of the $\Ncal$-extended SDYM model will take values in the adjoint 
representation of the algebra~$u(n)$. 
In the concise form the linear system (\ref{ls}) is written as
\begin{equation}\label{ls3}
\zeta^{\da}(\nabla_{\da}^i + 2\th^{i\a} \nabla_{\a\da})\,\psi \=0
\qquad\Leftrightarrow\qquad
\zeta^{\da}(\tilde D_{\da}^i + \tilde\Acal_{\da}^i)\,\psi \=0
\qquad\Leftrightarrow\qquad
\zeta^{\da}\,\tilde\nabla_{\da}^i\,\psi \=0\ .
\end{equation} 

\vspace{5mm} 
 
\noindent
{\bf 2.3 \ Reduction of $\Ncal$-extended SDYM to 2+1 dimensions} 
 
\noindent
The supersymmetric Bogomolny-type Yang-Mills-Higgs equations in 2+1 dimensions 
are obtained from the described $\Ncal$-extended super SDYM equations by a
dimensional reduction $\R^{2,2}\to\R^{2,1}$. In particular, for the $\Ncal{=}0$ 
sector we demand the components $A_\mu$ of a gauge potential to be independent 
of $x^4$ and put $A_4=:\vp$. Here, $\vp$ is a Lie-algebra valued scalar field 
in three dimensions (the Higgs field) which enters into the Bogomolny-type
equations. Similarly, for $\Ncal\ge 1$ one can reduce the $\Ncal$-extended 
SDYM equations on $\R^{2,2}$ by imposing the $\pa_4$-invariance condition on 
all the fields $(A_{\a\da},\ \c^i_{\a},\ \p^{[ij]},\ \tilde\c^{[ijk]}_{\da},
\ G_{\da\db}^{[ijkl]})$ from the $\Ncal{=}4$ supermultiplet or its truncation 
to $\Ncal{<}4$ and obtain supersymmetric Bogomolny-type equations on 
$\R^{2,1}$.

\smallskip

\noindent
{\bf Spinors in $\R^{2,1}$. } 
Recall that on $\R^{2,2}$ both $\Ncal{=}4$ SDYM theory and full $\Ncal{=}4$ 
super Yang-Mills theory have an SL(4, $\R$) $\cong$ Spin(3,3) R-symmetry 
group~\cite{Sieg}. A dimensional reduction to $\R^{2,1}$ enlarges the 
supersymmetry and R-symmetry to $2\Ncal{=}8$ and Spin(4,4), respectively, 
for both theories (cf.~\cite{Seib} for Minkowski signature). More generally,
any number $\Ncal$ of supersymmetries gets doubled to $2\Ncal$ in the 
reduction. Since dimensional reduction collapses the rotation group 
Spin(2,2) $\cong$ Spin(2,1)$_L\times$Spin(2,1)$_R$ of $\R^{2,2}$ to its 
diagonal subgroup Spin(2,1)$_D$ as the local rotation group of $\R^{2,1}$, 
the distinction between undotted and dotted indices disappears.
We shall use undotted indices henceforth.

\smallskip

\noindent
{\bf Coordinates and derivatives in $\R^{2,1}$. } The above discussion implies 
that one can relabel the bosonic coordinates $x^{\a\db}$ from (\ref{iso}) by 
$x^{\a\b}$ and split them as
\begin{equation}\label{iso2}
x^{\a\b}\=\sfrac12(x^{\a\b}+x^{\b\a})+\sfrac12(x^{\a\b}-x^{\b\a})\=
x^{(\a\b)}+x^{[\a\b]}
\end{equation}
into antisymmetric and symmetric parts,
\begin{equation}\label{iso3}
x^{[\a\b]}\= \sfrac12\ve^{\a\b}x^4 \= \sfrac12\ve^{\a\b}\tilde t \und
x^{(\a\b)}\ =:\ y^{\a\b} \ ,
\end{equation}
respectively, with
\begin{equation}\label{iso4}
y^{11}=x^{11}=\sfrac12(t-y)\ ,\qquad
y^{12}=\sfrac12(x^{12}+x^{21})=\sfrac12 x\ ,\qquad  
y^{22}=x^{22}=\sfrac12(t+y) \ .
\end{equation}
We also have $\th^{i\a}\mapsto\th^{i\a}$ and $\h^{\da}_i\mapsto\h^{\a}_i$
for the fermionic coordinates on $\R^{4|4\Ncal}$ reduced to~$\R^{3|4\Ncal}$.

Bosonic coordinate derivatives reduce in 2+1 dimensions to the operators
\begin{equation}\label{pdadb}
\pa_{(\a\b)}\=\sfrac12 (\pa_{\a\b}+\pa_{\b\a})
\end{equation}
which read explicitly as
\begin{equation}\label{p1212}
\pa_{(11)}\=\sfrac{\pa}{\pa y^{11}}\=\pa_{t}-\pa_y\ ,\qquad 
\pa_{(12)}=\pa_{(21)}\=\sfrac{1}{2}\sfrac{\pa}{\pa y^{12}}\=\pa_x\ ,\qquad
\pa_{(22)}\=\sfrac{\pa}{\pa y^{22}}\=\pa_{t}+\pa_y\  .
\end{equation}
We thus have  
\begin{equation}\label{pxdadb}
\frac{\pa}{\pa x^{\a\b}}\=\pa_{(\a\b)}-\ve_{\a\b}\pa_4\=
\pa_{(\a\b)}-\ve_{\a\b}\pa_{\tilde t} \ ,
\end{equation}
where $\ve_{12}=-\ve_{21}=-1$, $\pa_4={\pa}/{\pa x^{4}}$ and 
$\pa_{\tilde t}={\pa}/{\pa \tilde t}$.

The operators $D_{i\a}$ and $D^i_{\da}$ acting on $\tilde t$-independent 
superfields reduce to
\begin{equation}\label{hatD}
 D_{i\a} \=\pa_{i\a}+\h_i^{\b}\pa_{(\a\b )} \und
 D_{\a}^i\=\pa_{\a}^i+\th^{i\b}\pa_{(\a\b )}\ ,
\end{equation}
where $\pa_{i\a}={\pa}/{\pa\th^{i\a}}$ and $\pa_{\a}^i={\pa}/{\pa\h^{\a}_i}$.
Similarly, the antichiral operators $\tilde D_{i\a}$ and $\tilde D^i_{\da}$ 
in (\ref{pDD}) become 
\begin{equation}\label{hatDs}
\hat D_{i\a} \=\pa_{i\a} \und
\hat D_{\a}^i\=\pa_{\a}^i+2\th^{i\b}\pa_{(\a\b )}\ .
\end{equation}

\smallskip

\noindent
{\bf Supersymmetric Bogomolny-type equations in component fields. }
According to (\ref{pxdadb}), the components $A_{\a\db}$ of a gauge potential 
in four dimensions split into the components $A_{(\a\b)}$
of a gauge potential in three dimensions and a Higgs field 
$A_{[\a\b]}=-\ve_{\a\b}\,\vp$, i.e.
\begin{equation}\label{Adadb}
A_{\a\b} \= A_{(\a\b )} + A_{[\a\b ]} \=
A_{(\a\b )}-\ve_{\a\b}\,\vp\ .
\end{equation}
Then the covariant derivatives $D_{\a\db}$ reduced to three dimensions 
become the differential operators
\begin{equation}\label{2.42}
D_{\a\b} -\ve_{\a\b}\,\vp \= \pa_{(\a\b )} + [A_{(\a\b )},\ \cdot\ ]
-\ve_{\a\b}[\vp ,\ \cdot\ ] \ ,
\end{equation}
and the Yang-Mills field strength on $\R^{2,1}$ decomposes as
\begin{equation}\label{2.44}
F_{\a\b,\,\g\de} \= [D_{{\a}{\b}}, \, D_{\g\de}]
\= \ve_{\a\g}\,f_{\b\de}+\ve_{\b\de}\,f_{\a\g}
\qquad\textrm{with}\quad f_{\a\b}=f_{\b\a}\ .
\end{equation}

Substituting (\ref{Adadb}) and (\ref{2.42}) into (\ref{SDYMeom2}), i.e.\
demanding that all fields in (\ref{SDYMeom2}) are independent of 
$x^4=\tilde t$, we obtain the following supersymmetric Bogomolny-type 
equations on $\R^{2,1}$:
\begin{subequations}\label{2.43}
\begin{eqnarray}
&f_{\a\b}+D_{\a\b}\vp \=0\ ,
\\[2pt]
&D_{\a\b}\,\chi^{i\b} + \ve_{\a\b}\,[\vp ,\, \chi^{i\b} ]\=0\ ,
\\[2pt]
&D_{\a\b}\,D^{\a\b}\phi^{ij} + 2[\vp ,\, [\vp,\phi^{ij}]]+
2\{\chi^{i\a},\,\chi^j_{\a}\}\=0\ ,
\\[2pt]
&D_{\a\b}\,\tilde{\chi}^{\b[ijk]}-\ve_{\a\b}\,[\vp,\,
\tilde{\chi}^{\b[ijk]} ]-6[\chi_{\a}^{[i},\ \phi^{jk]} ]\=0\ ,
\\[2pt]
&D_{\a}^{\ \g} G_{\g\b}^{[ijkl]}+
[\vp,G_{\a\b}^{[ijkl]}]+
12\{\chi_{\a}^{[i},\tilde{\chi}_{\b}^{jkl]}\}-
18[\phi^{[ij},D_{\a\b}\phi^{{kl}]}]-
18\ve_{\a\b} [\phi^{[ij},[\phi^{{kl}]},\vp ]]\=0\ .
\end{eqnarray}
\end{subequations}

\smallskip

\noindent
{\bf Supersymmetric Bogomolny-type equations in terms of superfields. }
Translations generated by the vector field $\pa_4 = \pa_{\tilde t}$ are
isometries of superspaces $\R^{4|4\Ncal}$ and $\R^{4|2\Ncal}$. 
By taking the quotient with respect to the action of the abelian group~$\cal G$
generated by $\pa_4$, we obtain the reduced full superspace
$\R^{3|4\Ncal}\cong\R^{4|4\Ncal}/\cal G$ and the reduced antichiral superspace 
$\R^{3|2\Ncal}\cong\R^{4|2\Ncal}/\cal G$. In the following, we shall work
on $\R^{3|2\Ncal}$ and $\R^{3|2\Ncal}\times\C P^1$, since the reduced
$\psi$-function from (\ref{ls}) and (\ref{ls3}) is defined on the latter space.

The linear system stays in the center of the superfield approach to the 
$\Ncal$-extended SDYM equations. After imposing $\tilde t$-independence on all
fields in the linear system (\ref{ls3}), we arrive at the linear equations
\begin{equation}\label{ls5}
\zeta^{\a}\,\hat\nabla_{\a}^i\,\psi \ \equiv\
\zeta^{\a}(\hat D_{\a}^i + \hat\Acal_{\a}^i)\,\psi \=0
\end{equation} 
of the same form but with
\begin{equation}\label{2.46}
\hat D^i_{\a}\= \pa^i_{\a} + 2\th^{i\b}\pa_{(\a\b)} \und
\hat\Acal^i_{\a}\= \Acal^i_{\a} + 
2\th^{i\b}(\Acal_{(\a\b)}-\ve_{\a\b}\Xi )\ ,
\end{equation}
where $\Acal^i_{\a}$, $\Acal_{(\a\b)}$ and $\Xi$ are superfields 
depending on $y^{\a\b}$ and $\h_i^{\a}$ only.
These linear equations expand again to the pair (cf.~(\ref{ls}))
\begin{equation}\label{ls6}
\zeta^{\b}(\pa_{(\a\b)} + \Acal_{(\a\b)}- \ve_{\a\b}\Xi)\,\psi \=0 \und
\zeta^{\a}(\pa_{\a}^i + \Acal_{\a}^i)\,\psi \=0\ .
\end{equation}
The compatibility conditions for the linear system (\ref{ls5}) read
\begin{equation}\label{2.47}
\{\hat\nabla^i_{\a}, \hat\nabla^j_{\b}\}\ +\
\{\hat\nabla^i_{\b}, \hat\nabla^j_{\a}\}\=0
\qquad\Leftrightarrow\qquad
\{\hat\nabla^i_{\a},\hat\nabla^j_{\b}\}\=\ve_{\a\b}\,\hat\Fcal^{ij}
\end{equation}
and present a condensed form of (\ref{2.43}) rewritten in
terms of $\R^{3|2\Ncal}$ superfields.  
Similarly, these equations can also be written in more expanded forms 
analogously to (\ref{sdym1}) or using the superfield analog of (\ref{2.42}). 
However, we will not do this since all these sets of equations are equivalent.

\vspace{5mm} 
 
\section{Noncommutative $\Ncal$-extended U($n$) chiral model in 2+1 dimensions} 
 
\noindent
As has been known for some time, nonlinear sigma models in $2+1$ dimensions 
may be Lorentz-invariant or integrable but not both~\cite{Ward88, Zakr}. 
We will show that the super Bogomolny-type model discussed in Section 2 after 
a gauge fixing is equivalent to a super extension of the modified U($n$) chiral
model (so as to be integrable) first formulated by Ward~\cite{Ward88}. 
Since integrability is compatible with noncommutative deformation 
(if introduced properly, see e.g.~\cite{group5}--\cite{Penati}) 
we choose from the beginning to formulate our
super extension of this chiral model on Moyal-deformed $\R^{2,1}$ with
noncommutativity parameter~$\th\ge0$. Ordinary space-time  $\R^{2,1}$ can 
always be restored by taking the commutative limit $\th\to 0$.

\smallskip

\noindent
{\bf Star-product formulation. } Classical field theory on noncommutative 
spaces may be realized in a star-product formulation or in an operator 
formalism\footnote{See~\cite{noncom} for reviews on noncommutative field
theories.}. The first approach is closer to the commutative field theory: 
it is obtained by simply deforming the ordinary product of
classical fields (or their components) to the noncommutative star product
\begin{equation}\label{3.1}
(f\star g)(x) \= f(x) \exp\{\sfrac{\im}{2}\ \overleftarrow{\pa_a}\ 
\th^{ab}\ \overrightarrow{\pa_b}\}\,g(x)
\qquad\Rightarrow\qquad
x^a\star x^b - x^b\star x^a \=\im\th^{ab}
\end{equation}
with a constant antisymmetric tensor $\th^{ab}$. Specializing to $\R^{2,1}$, 
we use real coordinates  $(x^a)=(t,x,y)$ in which the Minkowski metric $g$ on 
$\R^3$ reads $(g_{ab})=\textrm{diag}(-1, +1,+1)$ with $a,b,\ldots=1,2,3$ 
(cf.~Section 2). 
It is straightforward to generalize the Moyal deformation (\ref{3.1})
to the superspaces introduced in the previous section, allowing in particular
for non-anticommuting Grassmann-odd coordinates. Deferring general superspace
deformations and their consequences to future work, we here content ourselves
with the simple embedding of the ``bosonic'' Moyal deformation into superspace,
meaning that (\ref{3.1}) is also valid for superfields $f$ and $g$ depending on 
Grassmann variables $\th^{i\a}$ and $\h^{\a}_i$. 

For later use we consider not only isotropic coordinates and vector fields
\begin{equation}\label{3.2}
u:=\sfrac12(t{+}y)=y^{22}\ ,\ \quad 
v:=\sfrac12(t{-}y)=y^{11}\ ,\ \quad 
\pa_u =\pa_t + \pa_y=\pa_{(22)}\ ,\ \quad 
\pa_v =\pa_t - \pa_y=\pa_{(11)}
\end{equation}
introduced in Section 2, but also the complex combinations
\begin{equation}\label{3.3}
z:=x+\im y\ ,\qquad 
\bar z:=x-\im y\ ,\qquad 
\pa_z =\sfrac12(\pa_x - \im\pa_y)\ ,\qquad
\pa_{\bar z} =\sfrac12(\pa_x +\im \pa_y )\ .
\end{equation}
Since the time coordinate $t$ remains commutative, 
the only nonvanishing component of the noncommutativity tensor $\th^{ab}$ is 
\begin{equation}\label{3.5}
\th^{xy} \= - \th^{yx}\ =:\ \th > 0 \qquad\Rightarrow\qquad 
\th^{z\bar z} \= - \th^{\bar z z} \= -2\im\,\th\ .
\end{equation}
Hence, we have
\begin{equation}\label{3.6}
z\star \bar z \= z\bar z +\th \und \bar z\star z \= z\bar z -\th
\end{equation} 
as examples of the general formula (\ref{3.1}).

\smallskip

\noindent
{\bf Operator formalism. } The nonlocality of the star products renders 
explicit computation cumbersome. We therefore pass to the operator formalism, 
which trades the star product for operator-valued spatial coordinates 
$(\hat x,\hat y)$ or their complex combinations $(\hat z,\hat{\bar z})$, 
subject to
\begin{equation}\label{3.7}
[t,\hat x]\=[t,\hat y]\=0 \quad\textrm{but}\quad [\hat x,\hat y]\=\im\th
\qquad\Rightarrow\qquad [\hat z,\hat{\bar z}]\=2\,\th\ .
\end{equation} 
The latter equation suggests the introduction of annihilation and creation 
operators,
\begin{equation}\label{3.8}
a\=\frac{1}{\sqrt{2\th}}\ \hat z \und
a^\+\=\frac{1}{\sqrt{2\th}}\ \hat{\bar z}
\qquad\textrm{with}\quad [a\,,a^\+]\=1\ ,
\end{equation} 
which act on a harmonic-oscillator Fock space $\Hcal$ with an orthonormal basis
$\{\,|\ell\>,\ \ell=0,1,2,\ldots\}$ such that
\begin{equation}\label{3.9}
a\,|\ell\> \= \sqrt{\ell}\;|\ell{-}1\> \und 
a^\+\,|\ell\> \= \sqrt{\ell{+}1}\;|\ell{+}1\>\ .
\end{equation} 

Any superfield $f(t,z,\bar z,\h^{\a}_i)$ on $\R^{3|2\Ncal}$
can be related to an operator-valued superfield 
$\hat f(t,\h^{\a}_i)\equiv F(t,a,a^\+,\h^{\a}_i)$ on $\R^{1|2\Ncal}$ 
acting in $\Hcal$, with the help of the Moyal-Weyl map
\begin{equation}\label{3.10}
f(t,z,\bar z,\h^{\a}_i)\quad\mapsto\quad \hat f(t,\h^{\a}_i)\= 
\textrm{Weyl-ordered}\ f\bigl(t,\sqrt{2\th}a,\sqrt{2\th}a^\+,\h^{\a}_i\bigr)\ .
\end{equation}
The inverse transformation recovers the ordinary superfield,
\begin{equation}\label{3.11}
\hat f(t,\h^{\a}_i)\ \equiv\
F(t,a,a^\+,\h^{\a}_i)\quad\mapsto\quad f(t,z,\bar z,\h^{\a}_i) \= F_{\star}
\bigl(t,\sfrac{z}{\sqrt{2\th}},\sfrac{\bar z}{\sqrt{2\th}},\h^{\a}_i\bigr)\ ,
\end{equation}
where $F_\star$ is obtained from $F$ by replacing ordinary with star products.
Under the Moyal-Weyl map, we have
\begin{equation}\label{3.12}
f\star g \quad\mapsto\quad \hat f\ \hat g \und
{\textstyle\int}\diff x\ \diff y\ f \= 2\pi\,\th\,\textrm{Tr} \hat f 
\=2\pi\,\th\sum_{\ell\ge 0}\<\ell |\hat f|\ell\> \ ,
\end{equation}
and the spatial derivatives are mapped into commutators,
\begin{equation}\label{3.13}
\pa_zf \quad\mapsto\quad \hat\pa_z\hat f \=
-\sfrac{1}{\sqrt{2\th}}\,[a^\+,\hat f] \und
\pa_{\bar z}f \quad\mapsto\quad \hat\pa_{\bar z}\hat f \=
\sfrac{1}{\sqrt{2\th}}\,[a\,,\hat f]\ .
\end{equation}
For notational simplicity we will from now on omit the hats over the operators
except when confusion may arise.

\smallskip

\noindent
{\bf Gauge fixing for $\psi$. }
Note that the linear system (\ref{ls5}) and the compatibility conditions 
(\ref{2.47}) are invariant under a gauge transformation
\begin{subequations}\label{gt}
\begin{eqnarray}
\j&\quad\mapsto\quad & \j'\=g^{-1}\j\ ,\\[2pt]
\Acal&\quad\mapsto\quad & \Acal' \=g^{-1}\Acal\,g + g^{-1}\pa\,g
\qquad\textrm{(with appropriate indices)}\ ,\\[2pt]
\Xi&\quad\mapsto\quad & \Xi'\= g^{-1}\Xi\,g\ ,
\end{eqnarray}
\end{subequations}
where $g=g(x^a, \h^{\a}_i)$ is a U($n$)-valued superfield globally defined on 
the deformed superspace $\R_\th^{3|2\Ncal}\times\C P^1$. 
Using a gauge transformation of the form (\ref{gt}), 
we can choose $\j$ such that it will satisfy the standard asymptotic 
conditions
(see e.g.~\cite{FT})
\begin{subequations}\label{ascon}
\begin{eqnarray}
\j&=&\Phi^{-1}\ +\ O(\zeta )\qquad\qquad\quad\,
\textrm{for}\quad\zeta\to 0\ ,\\[2pt]
\j&=&\unity\ +\ \zeta^{-1}\Ups\ +\ O(\zeta^{-2} )\quad
\textrm{for}\quad\zeta\to\infty\ ,
\end{eqnarray}
\end{subequations}
where the U($n$)-valued function $\Phi$ and $u(n)$-valued function $\Ups$ 
depend on $x^a$ and $\h_i^{\a}$. 
This ``unitary" gauge is compatible with the reality condition for $\j$,
\begin{equation}\label{3.19}
\j(x^a,\h_i^{\a},\zeta)\ \bigl[\j(x^a,\h_i^{\a},\bar\zeta)\bigr]^\+ \=\unity\ ,
\end{equation}
obtained by reduction from (\ref{recon}).

\smallskip

\noindent
{\bf Gauge fixing for $\hat\Acal^i_{\a}$. }
After fixing the unitary gauge (\ref{ascon}) for $\j$ and inserting
$(\z^{\a})=\bigl(\begin{smallmatrix}\z\\ -1\end{smallmatrix}\bigr)$ 
in the linear system~(\ref{ls5}), 
one can easily reconstruct the superfield given in~(\ref{2.46}) 
from $\Phi$ or $\Ups$ via
\begin{equation}\label{3.20}
\hat\Acal^i_{1}\=0 \und 
\hat\Acal^i_{2}\=\Phi^{-1}\hat D^i_{2}\Phi\=\hat D^i_{1}\Ups
\end{equation}
and thus fix a gauge for the superfields~$\hat\Acal^i_{\a}$. 
The operators $\hat D^i_{\a}$ were defined in~(\ref{hatDs}). 
One can express (\ref{3.20}) in terms of $\Acal^i_{\a}$ and 
$\Acal_{(\a\b)}-\ve_{\a\b}\Xi$ as
\begin{eqnarray}\label{3.17}
&\quad\Acal^i_1\=0 \und\qquad\quad 
\Acal^i_2\=\Phi^{-1}\pa^i_2\Phi\=\pa^i_1\Ups\ ,\\[2pt]
&\qquad\Acal_{(11)}\=0 \und 
\Acal_{(12)}+\Xi\=\Phi^{-1}\pa_{(12)}\Phi\= \pa_{(11)}\Ups\ ,\\[2pt]
&\Acal_{(21)}-\Xi\=0 \und\qquad
\Acal_{(22)}\=\Phi^{-1}\pa_{(22)}\Phi\= \pa_{(12)}\Ups\ .
\end{eqnarray}
Using (\ref{p1212}), we can rewrite the nonzero components as
\begin{equation}\label{3.24}
\Acal\ :=\ \Phi^{-1}\pa_{u}\Phi \= \pa_{x}\Ups\ ,\qquad
{\cal B}\ :=\ \Phi^{-1}\pa_{x}\Phi \= \pa_{v}\Ups\ ,\qquad
{\cal C}^i\ :=\ \Phi^{-1}\pa^i_2\Phi \= \pa^i_1\Ups\ .
\end{equation}
Recall that the superfields $\Phi$ and $\Ups$ depend on $x^a$ and $\h_i^{\a}$.

\smallskip

\noindent
{\bf Linear system. } 
In the above-introduced unitary gauge the linear system (\ref{ls6}) reads
\begin{equation}\label{3.25}
(\zeta\pa_x - \pa_u - \Acal )\,\j \= 0\ ,\qquad
(\zeta\pa_v - \pa_x - {\cal B})\,\j \= 0\ ,\qquad
(\zeta\pa_{1}^i - \pa_{2}^i - {\cal C}^i )\,\j \= 0\ ,
\end{equation}
which adds the last equation to the linear system of the Ward 
model~\cite{Ward88} and generalizes it to superfields $\Acal(x^a,\h_j^{\a})$, 
${\cal B}(x^a,\h_j^{\a})$ and ${\cal C}^i(x^a,\h_j^{\a})$.
The concise form of (\ref{3.25}) reads
\begin{equation}\label{3.28}
\bigl(\zeta\,\hat D_{1}^i - \hat D_{2}^i - \hat\Acal_{2}^i \bigr)\,\j \= 0
\end{equation}
or, in more explicit form,
\begin{equation}\label{3.29}
\Bigl[\zeta\,\bigl(\pa_1^i+2\th^{i1}\pa_v+2\th^{i2}\pa_x\bigr)\ -\
\bigl(\pa_2^i+{\cal C}^i+2\th^{i1}(\pa_x+{\cal B})+2\th^{i2}(\pa_u+\Acal)\bigr)
\Bigr]\,\j \=0\ .
\end{equation}

\smallskip

\noindent
{\bf $\Ncal$-extended sigma model. } The compatibility conditions of this 
linear system are the $\Ncal$-extended noncommutative sigma model equations
\begin{equation}\label{3.40}
\hat D_1^i(\Phi^{-1}\hat D_2^j\,\Phi )\ +\ 
\hat D_1^j(\Phi^{-1}\hat D_2^i\,\Phi )\=0
\end{equation}
which in expanded form reads
\begin{subequations}\label{3.41}
\begin{eqnarray}
&(g^{ab}+v_c\ve^{cab})\,\pa_a(\Phi^{-1}\pa_b\Phi)\=0\qquad\Leftrightarrow\qquad
\pa_x(\Phi^{-1}\pa_x\Phi)\ -\ \pa_v(\Phi^{-1}\pa_u\Phi)\=0\ ,\\[2pt]
&\pa_1^i(\Phi^{-1}\pa_x\Phi )\ -\ \pa_v(\Phi^{-1}\pa_2^i\Phi)\=0\ ,\qquad
\pa_1^i(\Phi^{-1}\pa_u\Phi)\ -\ \pa_x(\Phi^{-1}\pa_2^i\Phi)\=0\ ,\\[2pt]
&\pa_1^i(\Phi^{-1}\pa_2^j\Phi)\ +\ \pa_1^j(\Phi^{-1}\pa_2^i\Phi)\=0 \ .
\end{eqnarray}
\end{subequations}
Here, the first line contains the Wess-Zumino-Witten term with a
constant vector $(v_c)=(0,1,0)$ which spoils the standard Lorentz invariance 
but yields an integrable chiral model in 2+1 dimensions. Recall that $\Phi$ is 
a U($n$)-valued matrix whose elements act as operators in the Fock space 
$\Hcal$ and depend on $x^a$ and $2\Ncal$ Grassmann variables $\h_i^{\a}$. 
As discussed in Section 2, the compatibility conditions of the 
linear equations (\ref{3.28}) (or (\ref{3.25})) are equivalent to the 
$\Ncal$-extended Bogomolny-type equations (\ref{2.43}) for the component
(physical) fields. Thus, chiral model field equations (\ref{3.41}) are 
equivalent to a gauge fixed form of equations (\ref{2.43}).

\smallskip

\noindent
{\bf $\Ups$-formulation. } Instead of $\Phi$-parametrization of 
$(\Acal,{\cal B},{\cal C}^i)$ given in (\ref{3.17})--(\ref{3.24}) 
we may use the equivalent $\Ups$-parametrization also given there.
In this case, the compatibility conditions for the linear system~(\ref{3.25}) 
reduce to 
\begin{subequations}\label{3.44}
\begin{eqnarray}
&(\pa_x^2 - \pa_u\pa_v)\Ups\ +\ [\pa_v\Ups\,,\,\pa_x\Ups]\=0\ , \\[2pt]
&(\pa_2^i\pa_v- \pa^i_1\pa_x)\Ups\ +\ [\pa^i_1\Ups\,,\,\pa_v\Ups]\=0\ ,\qquad
 (\pa_2^i\pa_x- \pa^i_1\pa_u)\Ups\ +\ [\pa^i_1\Ups\,,\,\pa_x\Ups]\=0\ ,\\[2pt]
&(\pa_2^i\pa_1^j+\pa_2^j\pa^i_1)\Ups\ +\ \{\pa^i_1\Ups\,,\,\pa_1^j\Ups\}\=0\ ,
\end{eqnarray}
\end{subequations}
which in concise form read
\begin{equation}\label{3.47}
(\hat D_2^i\,\hat D_1^j + \hat D_2^j\,\hat D^i_1)\,\Ups\ +\ 
\{\hat D^i_1\Ups\,,\,\hat D_1^j\Ups\} \=0\ . 
\end{equation}
Recall that $\Ups$ is a $u(n)$-valued matrix whose elements act as operators 
in the Fock space~$\Hcal$ and depend on $x^a$ and $2\Ncal$ Grassmann 
variables~$\h_i^{\a}$.

For ${\Ncal}{=}4$, the commutative limit of (\ref{3.47}) can be considered as
Siegel's equation~\cite{Sieg} reduced to 2+1 dimensions. 
According to Siegel, one can extract the multiplet of physical fields 
appearing in (\ref{2.43}) from the prepotential $\Ups$ via 
\begin{subequations}\label{3.48}
\begin{eqnarray}
&\pa_1^i\Ups = A_2^i\ ,\qquad
 \pa_1^i\pa_1^j\Ups = \p^{ij}\ ,\qquad
 \pa_1^i\pa_1^j\pa_1^k\Ups = \tilde\chi^{[ijk]}_2\ ,\qquad
 \pa_1^i\pa_1^j\pa_1^k\pa_1^l\Ups = G^{[ijkl]}_{22}\ ,\\[2pt]
&\pa_{(\a1)}\Ups = A_{(\a2)} -\ve_{\a2}\vp\ ,\qquad
 \pa_{(\a1)}\pa_1^i\Ups =\chi^i_{\a}\ ,\qquad
 \pa_{(\a1)}\pa_{(\b1)}\Ups = f_{\a\b}\ ,
\end{eqnarray}
\end{subequations}
where one takes $\Ups$ and its derivatives at $\h_i^2=0$. 
The other components of the physical fields, i.e.~$\tilde\chi^{[ijk]}_1$, 
$G^{[ijkl]}_{11}$, $G^{[ijkl]}_{21}$, $A_{(11)}$ and $A_{(21)}{-}\vp$, 
vanish in this {\it light-cone\/} gauge.

\smallskip

\noindent
{\bf Supersymmetry transformations. } The $4\Ncal$ supercharges given in 
(\ref{2.12}) reduce in 2+1 dimensions to the form
\begin{equation}\label{1}
Q_{i\a}\=\pa_{i\a}-\h_i^{\b}\pa_{(\a\b)} \und
Q^i_{\a}\=\pa^i_{\a}-\th^{i\b}\pa_{(\a\b)}\ .
\end{equation}
Their antichiral version, matching to $\hat D_{i\a}$ and $\hat D^j_{\b}$ 
of~(\ref{hatDs}), reads
\begin{equation}\label{2}
\hat Q_{i\a}\=\pa_{i\a}-2\h_i^{\b}\pa_{(\a\b)} \und
\hat Q^j_{\b}\=\pa^j_{\b}\ ,
\end{equation}
so that
\begin{equation}\label{3}
\{\hat Q_{i\a}\,,\,\hat Q^j_{\b}\}\=-2\,\de^j_i\,\pa_{(\a\b)}\ .
\end{equation}

On a (scalar) $\R^{3|2\Ncal}$ superfield $\Sigma$ these supersymmetry 
transformations act as
\begin{equation}\label{5}
\hat\de\,\Sigma\ :=\ 
\ve^{i\a}\hat Q_{i\a}\Sigma\ +\ \ve_i^{\a}\hat Q^i_{\a}\Sigma
\end{equation}
and are induced by the coordinate shifts
\begin{equation}\label{4}
\hat\de\,y^{\a\b} \= -2\ve^{i(\a}\h_i^{\b)} \und
\hat\de\,\h_i^{\a}\= \ve_i^{\a}\ ,
\end{equation}
where $\ve^{i\a}$ and $\ve_i^{\a}$ are $4\Ncal$ real Grassmann parameters.
It is easy to see that our equations (\ref{3.40}) and (\ref{3.47}) are 
invariant under the supersymmetry transformations (\ref{5}) 
(applied to $\Phi$ or $\Ups$). This is simply because the operators 
$\hat D_{i\a}$ and $\hat D^j_{\b}$ anticommute with the supersymmetry 
generators $\hat Q_{i\a}$ and $\hat Q^j_{\b}$. 
Therefore, the equations of motion (\ref{3.41})
of the modified $\Ncal$-extended chiral model in 2+1 dimensions as well as 
their reductions to 2+0 and 1+1 dimensions carry $2\Ncal$ supersymmetries and
are genuine supersymmetric extensions of the corresponding bosonic equations.
Note that this type of extension is not the standard one since 
the R-symmetry groups are Spin($\Ncal,\Ncal$) in 2+1 
and Spin($\Ncal,\Ncal$)$\times\,$Spin($\Ncal,\Ncal$) in 1+1 dimensions, which 
differ from the compact unitary R-symmetry groups of standard sigma models.
Contrary to the standard case of two-dimensional sigma models 
the above ``noncompact" $2\Ncal$ supersymmetries do not impose any constraints
on the geometry of the target space,
e.g.~they do not demand it to be K\"ahler~\cite{Zumino} or 
hyper-K\"ahler~\cite{AGF}. This may be of interest and deserves further study.

\smallskip

\noindent
{\bf Action functionals. } In either formulation of the $\Ncal$-extended 
supersymmetric SDYM model on $\R^{2,2}$ there are difficulties with finding 
a proper action functional generalizing the one~\cite{group10, New} 
for the purely bosonic case. 
These difficulties persist after the reduction to 2+1 dimensions, 
i.e.\ for the equations (\ref{3.41}) and (\ref{3.44}) describing our 
supersymmetric modified U($n$) chiral model. It is the price to be paid for 
overcoming the no-go barrier $\Ncal\le4$ and the absence of geometric 
target-space constraints. On a more formal level, the problem is related to 
the {\it chiral\/} character of (\ref{3.40}) as well as (\ref{3.47}), where 
only the operators $\hat D^i_{\a}$ but not $\hat D_{i\a}$ appear. 
Note however, that for $\Ncal=4$ one {\it can\/} write an
action functional in component fields producing the equations (\ref{2.43}), 
which are equivalent to the superspace equations (\ref{3.40}) 
when $i,j=1,\ldots,4$ (see e.g.~\cite{sigma8}).

One proposal for an action functional stems from Siegel's 
idea~\cite{Sieg} for the $\Ups$-formulation of the $\Ncal$-extended SDYM 
equations. Namely, one sees that $\pa^i_2\Ups$ enters only linearly into the 
last two lines in~(\ref{3.44}). Therefore, if we introduce 
\begin{equation}
\Ups_{(1)}\ :=\ \Ups|_{\h^2_i=0}
\end{equation}
then it must satisfy the first equation from (\ref{3.44}), and the remaining 
equations iteratively define the dependence of $\Ups$ on $\h_i^2$ starting from 
$\Ups_{(1)}$. Hence, all information is contained in $\Ups_{(1)}$, as can also 
be seen from~(\ref{3.48}). In other words, the dependence of $\Ups$ on $\h_i^2$ 
is not `dynamical'. For an action one can then take (cf.~\cite{Sieg})
\begin{equation}\label{9}
S\=\int\!\!\diff^3x\;\diff^{\Ncal}\h^{1}\ \bigl\{
\Ups_{(1)}\pa_{(\a\b)}\pa^{(\a\b)}\Ups_{(1)}\ +\ \sfrac23\, 
\Ups_{(1)}\,\ve^{\a\b}\pa_{(\a1)}\Ups_{(1)}\,\pa_{(\b1)}\Ups_{(1)}\bigr\}\ .
\end{equation} 
Extremizing this functional yields the first line of (\ref{3.44}) at $\h^2_i=0$.
Except for the Grassmann integration, this action has the same form as the
purely bosonic one~\cite{New}. One may apply the same logic to the 
$\Phi$-formulation where the action for the purely bosonic case is also 
known~\cite{group10, IoZak}.

\vspace{5mm} 
 
\section{$\Ncal$-extended multi-soliton configurations via dressing} 
 
\noindent
The existence of the linear system (\ref{3.28}) (equivalent to (\ref{3.25}))
encoding solutions of the $\Ncal$-extended U($n$) chiral model in an auxiliary
matrix $\j$ allows for powerful methods to systematically construct explicit
solutions for $\j$ and hence for $\ \Phi^\+=\j|_{\zeta=0}\ $ and 
$\ \Ups=\lim\limits_{\zeta\to\infty}\zeta\,(\j{-}\unity)$.
For our purposes the so-called dressing method~\cite{group11, FT} proves to be 
the most practical~\cite{LP1}--\cite{Penati}, and so we shall use it here for 
our linear system, i.e.\ already in the $\Ncal$-extended noncommutative case.

\smallskip

\noindent
{\bf Multi-pole ansatz for $\j$. } The dressing method is a recursive procedure
for generating a new solution from an old one. More concretely, we rewrite the 
linear system~(\ref{3.25}) in the form
\begin{equation}\label{4.1}
\j (\pa_u - \zeta\pa_x )\j^\+ \= \Acal \ ,\qquad
\j (\pa_x - \zeta\pa_v )\j^\+ \= {\cal B} \ ,\qquad
\j (\pa_2^i - \zeta\pa_1^i )\j^\+ \= {\cal C}^i \ .
\end{equation}
Recall that $\j^\+:=(\j(x^a,\h_i^{\a},\bar\zeta))^\+$ and 
$(\Acal,{\cal B},{\cal C}^i)$ depend only on $x^a$ and $\h_i^{\a}$. 
The central idea is to demand analyticity in the spectral parameter $\zeta$,
which strongly restricts the possible form of~$\j$. One way to exploit this
constraint starts from the observation that the left hand sides of (\ref{4.1}) 
as well as of the reality condition (\ref{3.19}) do not depend on $\zeta$
while $\j$ is expected to be a nontrivial function of~$\zeta$ globally defined 
on~$\C P^1$. Therefore, it must be a meromorphic function on~$\C P^1$ 
possessing some poles which we choose to lie at finite points with constant
coordinates $\mu_k\in\C P^1$.

Here we will build a (multi-soliton) solution $\j_m$ featuring $m$ simple 
poles at positions $\mu_1,\ldots,\mu_m$ with\footnote{
This condition singles out solitons over anti-solitons, 
which appear for Im$\,\mu_k>0$.}
Im$\,\mu_k<0$ by left-multiplying an $(m{-}1)$-pole solution $\j_{m-1}$ 
with a single-pole factor of the form
\begin{equation}\label{4.4}
\unity\ +\ \frac{\mu_m -\bar\mu_m}{\zeta -\mu_m} \ P_m(x^a,\h_i^{\a})\ , 
\end{equation}
where the $n{\times}n$ matrix function $P_m$ is yet to be determined.
Starting from the trivial (vacuum) solution $\j_0 =\unity$, the iteration
$\j_0\mapsto\j_1\mapsto\ldots\mapsto\j_m$ yields a multiplicative ansatz 
for~$\j_m$,
\begin{equation}\label{4.5}
\j_m\=\prod\limits^{m-1}_{\ell=0}\Bigl(\unity\ +\ 
\frac{\mu_{m-\ell}-\bar\mu_{m-\ell}}{\zeta-\mu_{m-\ell}}\ P_{m-\ell}\Bigr)\ , 
\end{equation}
which, via partial fraction decomposition, may be rewritten in the additive 
form 
\begin{equation}\label{4.6}
\j_m\=\unity\ +\ \sum\limits^{m}_{k=1}\frac{\La_{mk}S^\+_k}{\zeta -\mu_{k}}\ , 
\end{equation}
where $\La_{m k}$ and $S_k$ are some $n{\times}r_k$ matrices depending on 
$x^a$ and $\h_i^{\a}$, with $r_k\le n$.

\smallskip

\noindent
{\bf Equations for $S_k$. } Let us first consider the additive parametrization 
(\ref{4.6}) of $\j_m$. This ansatz must satisfy the reality condition 
(\ref{3.19}) as well as our linear equations in the form (\ref{4.1}). In
particular, the poles at $\zeta =\bar\mu_k$ on the left hand sides of these
equations have to be removable since the right hand sides are independent of 
$\zeta$. Inserting the ansatz (\ref{4.6}) and putting to zero the corresponding
residues, we learn from (\ref{3.19}) that
\begin{equation}\label{4.7}
\Bigl(\unity\ +\ \sum\limits^{m}_{\ell=1}
\frac{\La_{m\ell}S^\+_\ell}{\bar\mu_k - \mu_\ell}\Bigr)\,S_k \=0\ , 
\end{equation}
while from (\ref{4.1}) we obtain the differential equations
\begin{equation}\label{4.8}
\Bigl(\unity\ +\ \sum\limits^{m}_{\ell=1}
\frac{\La_{m\ell}S^\+_\ell}{\bar\mu_k-\m_\ell}\Bigr)\,
\bar L_k^{\Acal,{\cal B},i}\,S_k\=0\ ,
\end{equation}
where $\bar L_k^{\Acal,{\cal B},i}$ stands for either
\begin{equation}\label{4.9}
\bar L_k^{\Acal} \= \pa_u -\bar\mu_k\pa_x\ ,\qquad 
\bar L_k^{\cal B} \= \mu_k(\pa_x -\bar\mu_k\pa_v) \qquad\textrm{or}\qquad 
\bar L_k^{i} \= \pa^i_2 - \bar\mu_k\pa_1^i\ .
\end{equation}
Note that we consider a {\it recursive} procedure starting from $m{=}1$, 
and operators (\ref{4.9}) will appear with $k=1,\ldots ,m$ if we consider 
poles at $\zeta =\bar\mu_k$.

Because the $\bar L_k^{\Acal,{\cal B},i}$ for $k=1,\ldots,m$ are linear
differential operators, it is easy to write down the general solution for 
(\ref{4.8}) at any given~$k$, by passing from the coordinates 
$(u,v,x;\h_i^1,\h_i^2)$ to ``co-moving coordinates'' 
$(w_k,\bar w_k,s_k;\h^i_k,\bar\h^i_k)$. 
The precise relation for $k=1,\ldots,m$ is~\cite{LP1, ChuLe}
\begin{equation}\label{4.10}
w_k\ :=\ x + \bar\mu_k u + \bar\mu_k^{-1}v \= 
x+\sfrac12(\bar\mu_k{-}\bar\mu_k^{-1})y+\sfrac12(\bar\mu_k{+}\bar\mu_k^{-1})t
\und
\h^i_k\ :=\ \h_i^1+\bar\mu_k\h_i^2 \ ,
\end{equation}
with $\bar w_k$ and $\bar\h^i_k$ obtained by complex conjugation and the
co-moving time~$s_k$ being inessential because by definition nothing will 
depend on it. The $k$th moving frame travels with a constant velocity
\begin{equation}
(\textrm{v}_x\,,\,\textrm{v}_y)_k \= - \Bigl( 
\frac{\mu_k+\bar\mu_k}{\mu_k\bar\mu_k+1}\;,\,
\frac{\mu_k\bar\mu_k-1}{\mu_k\bar\mu_k+1} \Bigr) \ ,
\end{equation}
so that the static case $w_k{=}z$ is recovered for $\mu_k=-\im$.
On functions of $(w_k,\h^i_k,\bar w_k,\bar\h^i_k)$ alone the operators
(\ref{4.9}) act as
\begin{equation}\label{4.12}
\bar L_k^{\Acal} \= \bar L_k^{\cal B} \= 
(\mu_k{-}\bar\mu_k)\frac{\pa}{\pa\bar w_k}\ =:\ \bar L_k \und
\bar L_k^i \= (\mu_k{-}\bar\mu_k)\frac{\pa}{\pa\bar \h_k^i}\ .
\end{equation}
By induction in $k=1,\ldots ,m$ we learn that, due to (\ref{4.7}), a necessary
and sufficient condition for a solution of (\ref{4.8}) is
\begin{equation}\label{4.13}
\bar L_k S_k \= S_k \tilde Z_k \und
\bar L_k^{i} S_k \= S_k \tilde Z_k^i 
\end{equation}
with some $r_k{\times}r_k$ matrices $\tilde Z_k$ and $\tilde Z_k^i$
depending on $(w_k,\bar w_k,\h_k^j,\bar\h_k^j)$.

Passing to the noncommutative bosonic coordinates we obtain
\begin{equation}\label{4.14}
\bigl[\hat w_k\,,\,\hat{\bar w}_k\bigr]\= 2\th\,\nu_k\bar\nu_k 
\qquad\textrm{with}\qquad
\nu_k\bar\nu_k \= \sfrac{4\im}{\mu_k-\bar\mu_k-\mu_k^{-1}+\bar\mu_k^{-1}}\ .
\end{equation}
Thus, we can introduce annihilation and creation operators
\begin{equation}\label{4.15}
c_k\=\frac{1}{\sqrt{2\th}}\frac{\hat w_k}{\nu_k} \und
c_k^\+\=\frac{1}{\sqrt{2\th}}\frac{\hat{\bar w}_k}{\bar\nu_k}
\qquad\textrm{so that}\quad [c_k\,,\,c_k^\+]\=1
\end{equation}
for $k=1,\ldots,m$. 
Naturally, this Heisenberg algebra is realized on a ``co-moving'' Fock 
space~$\Hcal_k$, with basis states~$|\ell\>_k$ and a ``co-moving'' 
vacuum~$|0\>_k$ subject to $c_k|0\>_k=0$.
Each co-moving vacuum $|0\>_k$ (annihilated by $c_k$) is related to the 
static vacuum $|0\>$ (annihilated by $a$) through an ISU(1,1) squeezing
transformation (cf.~\cite{LP1}) which is time-dependent. 
The fermionic coordinates $\h^i_k$ and $\bar\h^i_k$ remain
spectators in the deformation. 
Coordinate derivatives are represented in the standard fashion as
\begin{equation}\label{4.16}
\nu_k\sqrt{2\th}\frac{\pa}{\pa w_k}\quad\mapsto\quad-[c_k^\+\,,\ \cdot\ ] \und
\bar\nu_k\sqrt{2\th}\frac\pa{\pa\bar w_k}\quad\mapsto\quad[c_k\,,\ \cdot\ ]\ .
\end{equation}

After the Moyal deformation, the $n{\times}r_k$ matrices $S_k$ have become
operator-valued, but are still functions of the Grassmann coordinates
$\h_k^i$ and~$\bar\h_k^i$. The noncommutative version of the
BPS conditions~(\ref{4.13}) naturally reads
\begin{equation}\label{SBPS}
c_k\,S_k \= S_k\,Z_k \und \frac{\pa}{\pa\bar\h_k^i}\,S_k \= S_k\,Z_k^i
\end{equation}
where $Z_k$ and $Z_k^i$ are some operator-valued $r_k{\times}r_k$ matrix
functions of  $\h_k^j$ and $\bar\h_k^j$.

\smallskip

\noindent
{\bf Nonabelian solutions for $S_k$. }
For general data $Z_k$ and $Z_k^i$ it is difficult to solve (\ref{SBPS}),
but it is also unnecessary because the final expression~$\psi_m$  
turns out not to depend on them. Therefore, we conveniently choose
\begin{equation}\label{Shol}
Z_k=c_k\otimes\unity_{r_k\times r_k} \quad\textrm{and}\quad Z_k^i=0
\qquad\Rightarrow\qquad S_k \= R_k(c_k,\h_k^i)\ ,
\end{equation}
where $R_k$ is an arbitrary $n{\times}r_k$ matrix function independent of 
$c_k^\+$ and $\bar\h_k^i$.\footnote{
Changing $Z_k$ or $Z_k^i$ multiplies $R_k$ by an invertible factor from the 
right, which drops out later, except for the degenerate case $Z_k{=}0$ 
which yields \ $S_k=R_k\,|0\>_k\<0|_k$.}
It is known that nonabelian (multi-) solitons arise for algebraic 
functions~$R_k$ (cf.~e.g.~\cite{Ward88} for the commutative
and~\cite{LP1} for the noncommutative $\Ncal{=}0$ case).
Their common feature is a smooth commutative limit. The only novelty
of the supersymmetric extension is the $\h_k^i$~dependence, i.e.
\begin{equation}\label{Rexpand}
R_k \=R_{k,0}+\h_k^i R_{k,i}+\h_k^i\h_k^j R_{k,ij}+
\h_k^i\h_k^j\h_k^p R_{k,ijp}+\h_k^i\h_k^j\h_k^p\h_k^q R_{k,ijpq}\ .
\end{equation}

\smallskip

\noindent
{\bf Abelian solutions for $S_k$. }
It is useful to view $S_k$ as a map from 
$\C^{r_k}\otimes\Hcal_k$ to $\C^n\otimes\Hcal_k$
(momentarily suppressing the $\h$~dependence).
The noncommutative setup now allows us to generalize
the domain of this map to {\it any\/} subspace of
$\C^n\otimes\Hcal_k$. In particular, we may choose it
to be finite-dimensional, say $\C^{q_k}$, and represent 
the map by an $n{\times}q_k$ array $|S_k\>$ of kets in~$\Hcal$.
In this situation, $Z_k$ and $Z_k^i$ in~(\ref{SBPS}) are just 
{\it number\/}-valued $q_k{\times}q_k$ matrix functions of $\h_k^j$
and~$\bar\h_k^j$. In case they do not depend on $\bar\h_k^j$,
we can write down the most general solution as
\begin{equation}\label{Sabelian}
|S_k\>\=R_k(c_k,\h_k^j)\ |Z_k\>\ 
\exp\bigl\{\textstyle{\sum_i} Z_k^i(\h_k^j)\,\bar\h_k^i\bigr\}
\qquad\textrm{with}\quad
 |Z_k\>\ :=\ \exp\bigl\{Z_k(\h_k^j)\,c_k^\+\bigr\}\,|0\>_k \ .
\end{equation}
As before, we may put $Z_k^i=0$ without loss of generality,
but now the choice of $Z_k$ does matter.

For any given~$k$ generically there exists a $q_k$-dimensional 
basis change which diagonalizes the ket-valued matrix
\begin{equation}\label{Zdiag}
|Z_k\>\quad\mapsto\quad \textrm{diag}\,
\bigl(\e^{\a_k^1 c^\+},\e^{\a_k^2 c^\+},\ldots,
\e^{\a_k^{q_k}c^\+}\bigr)\,|0\>_k \=\textrm{diag}\,
\bigl(|\a_k^1\>\,,\,|\a_k^2\>\,,\ldots,\,|\a_k^{q_k}\>\bigr)\ ,
\end{equation}
where we defined coherent states
\begin{equation}\label{coherent}
|\a_k^l\>\ :=\ \e^{\a_k^l c^\+}|0\>_k \qquad\textrm{so that}\qquad
c_k\,|\a_k^l\> \= \a_k^l\,|\a_k^l\> \qquad\textrm{for}\quad
l=1,\ldots,q_k\quad\textrm{and}\quad \a_k^l\in\C\ .
\end{equation}
Note that not only the entries of $R_k$ but also the $\a_k^l$ are 
holomorphic functions of the co-moving Grassmann parameters~$\h_k^j$ 
and thus can be expanded like in~(\ref{Rexpand}).
In the U(1) model, we must use ket-valued $1{\times}q_k$ matrices
$|S_k\>$ for all~$k$, yielding rows
\begin{equation}\label{Sn1}
|S_k\>\=\bigl( R_k^1\,|\a_k^1\>\,,\,R_k^2\,|\a_k^2\>\,,\ldots,\,
R_k^{q_k}\,|\a_k^{q_k}\>\bigr)
\qquad\textrm{for}\quad k=1,\ldots,m \ ,
\end{equation}
with functions $\a_k^l(\h_k^j)$. Here, the $R_k^l$ only affect the
states' normalization and can be collected in a diagonal matrix
to the right, hence will drop out later and thus may all be put to one.
Formally, we have recovered the known abelian (multi-) soliton solutions,
but the supersymmetric extension has generalized $|S_k\>\to|S_k(\h_k^j)\>$.

\smallskip

\noindent
{\bf Explicit form of $P_k$. } Let us now consider the multiplicative 
parametrization (\ref{4.5}) of $\j_m$ which also allows us to 
solve~(\ref{4.7}). First of all, note that the reality condition~(\ref{3.19})
is satisfied if
\begin{equation}\label{4.21}
P_k \= P_k^\+ \= P_k^2\qquad\Leftrightarrow\qquad 
P_k \= T_k\,(T_k^\+T_k)^{-1}T_k^\+ \qquad\textrm{for}\quad
k=1,\ldots,m\ ,
\end{equation}
meaning that $P_k$ is an operator-valued hermitian projector
(of group-space rank $r_k\le n$) built from an $n{\times}r_k$ 
matrix function $T_k$ (the abelian case of $n{=}1$ is included).
The reality condition follows just because
\begin{equation}\label{4.22}
\Bigl(\unity\ +\ \frac{\mu_k -\bar\mu_k}{\zeta - \mu_k}\ P_k\Bigr)
\Bigl(\unity\ +\ \frac{\bar\mu_k - \mu_k}{\zeta - \bar\mu_k}\ P_k\Bigr)\=\unity
\qquad\textrm{for any $\zeta$ and $k=1,\ldots ,m$}\ .
\end{equation}
The $r_k$ columns of $T_k$ span the image of $P_k$ and obey
\begin{equation}\label{4.23}
P_k\,T_k\=T_k \qquad\Leftrightarrow\qquad (\unity{-}P_k)\,T_k\=0\ .
\end{equation}
Furthermore, the equation (\ref{4.7}) with $m=k$ (induction) rewritten 
in the form
\begin{equation}\label{4.24}
(\unity{-}P_k)\ \prod\limits_{\ell=1}^{k-1}\Bigl(\unity\ +\ 
\frac{\mu_{k-\ell} -\bar\mu_{k-\ell}}{\bar\mu_k - \mu_{k -\ell}}\ 
P_{k-\ell}\Bigr)\ S_k \=0
\end{equation}
reveals that (cf.~(\ref{4.23}))
\begin{equation}\label{4.25}
T_1 \= S_1 \und
T_k \=\biggl\{\prod\limits_{\ell =1}^{k-1}\Bigl(\unity\ -\ 
\frac{\mu_{k-\ell} -\bar\mu_{k-\ell}}{\mu_{k-\ell} - \bar\mu_{k}}\ 
P_{k-\ell}\Bigr)\biggr\}\ S_k \qquad\textrm{for}\quad k\ge2\ ,
\end{equation}
where the explicit form of $S_k$ for $k=1,\ldots ,m$ 
is given in (\ref{Shol}) or (\ref{Sabelian}).
The final result reads 
\begin{equation}\label{4.26}
\j_m\=\prod\limits_{\ell=0}^{m-1}\Bigl(\unity\ +\ 
\frac{\mu_{m-\ell} -\bar\mu_{m-\ell}}{\zeta - \mu_{m-\ell}}\ P_{m-\ell}\Bigr) 
\= \unity\ +\ \sum\limits_{k=1}^{m}\frac{\La_{mk}S^\+_k}{\zeta - \mu_{k}}
\end{equation}
with hermitian projectors $P_k$ given by (\ref{4.21}), $T_k$ given 
by (\ref{4.25}) and $S_k$ given by (\ref{Shol}) or (\ref{Sabelian}). 
The explicit form of $\La_{mk}$ (which we do not need) can be found 
in~\cite{LP1}. The corresponding superfields $\Phi$ and $\Ups$ are
\begin{subequations}\label{4.27}
\begin{eqnarray}
&&\Phi_m \= \j^\+_m|_{\zeta=0} \= \prod\limits_{k=1}^{m}(\unity-\rho_kP_k) 
\qquad\textrm{with}\quad \rho_k\=1-\frac{\mu_k}{\bar\mu_k}\ ,\\
&&\Ups_m \= \lim\limits_{\zeta\to\infty}\zeta\,(\j_m -\unity)
\=\sum\limits_{k=1}^{m} (\mu_{k}{-}\bar\mu_{k})\,P_k\ .
\end{eqnarray}
\end{subequations} 

{}From (\ref{4.21}) it is obvious that $P_k$ is invariant under a similarity
transformation
\begin{equation}\label{4.28}
T_k\ \mapsto\ T_k\,\La_k \qquad\Leftrightarrow\qquad
S_k\ \mapsto\ S_k\,\La_k
\end{equation}
for an invertible operator-valued $r_k{\times}r_k$ matrix~$\La_k$.
This justifies putting $Z_k^i=0$ from the beginning and also the restriction  
to $Z_k=c_k\otimes\unity_{r_k\times r_k}$ in the nonabelian case,
both without loss of generality.
Hence, the nonabelian solution space constructed here is parametrized by
the set $\{R_k\}^m_1$ of matrix-valued functions of $c_k$ and $\h_k^i$
and the pole positions $\mu_k$.
The abelian moduli space, however,  is larger by the set $\{Z_k\}^m_1$ 
of matrix-values functions of~$\h_k^i$ which generically contain the 
coherent-state parameter functions~$\{\a_k^l(\h_k^i)\}$. 
Restricting to $\h_k^i{=}0$ reproduces the soliton configurations of
the bosonic model~\cite{LP1}.

\smallskip

\noindent
{\bf Static solutions. } Let us consider the reduction to 2+0 dimensions, 
i.e.~the static case. Recall that static solutions correspond to the choice 
$m=1$ and $\m_1\equiv\m =-\im$ implying $w_1=z$, so we drop the index~$k$.
Specializing (\ref{4.26}), we have 
\begin{equation}\label{4.29}
\j \= \unity\ -\ \frac{2\,\im}{\zeta +\im}\,P \qquad\textrm{so that}\qquad 
\Phi \= \Phi^\+ \= \unity-2P\ ,
\end{equation}
where a hermitian projector $P$ of group-space rank~$r$
satisfies the BPS equations
\begin{subequations}\label{4.30}
\begin{eqnarray} 
(\unity{-}P)\,a\,P\=0 & \Rightarrow & (\unity{-}P)\,a\,T\=0\ ,\\[2pt]
(\unity{-}P)\,\frac{\pa}{\pa\bar\h^i}P\=0 & \Rightarrow & 
(\unity{-}P)\,\frac{\pa}{\pa\bar\h^i}T\=0\ ,
\end{eqnarray}
\end{subequations} 
with $P=T\,(T^\+T)^{-1}T^\+$ and $\h^i=\h_i^1+\im\h_i^2$. In this case $T=S$, 
and for a nonabelian $r{=}1$ projector $P$ we get $T=T(a,\h^i)$ as an 
$n{\times}1$ column. For the simplest case of $\Ncal{=}1$ we just have 
(cf.~\cite{Perelomov})
\begin{equation}\label{4.31}
T\=T_e(a)\ +\ \h\,T_o(a) \qquad\textrm{with}\quad \h = \h^1 +\im\h^2\ ,
\end{equation}
where $T_e(a)$ and  $T_o(a)$ are rational functions of $a$ (e.g.~polynomials) 
taking values in the even and odd parts of the Grassmann algebra.
Similarly, an abelian $\Ncal{=}1$ projector (for $n{=}1$) is built from 
\begin{equation}\label{4.31a}
|T\>\=\bigl(\,|\a^1\>\,,\,|\a^2\>\,,\ldots,\,|\a^q\>\bigr)\ .
\end{equation}

At $\th{=}0$, 
the static solution (\ref{4.31}) of our supersymmetric U($n$) sigma model is 
also a solution of the standard $\Ncal{=}1$ supersymmetric $\C P^{n-1}$ sigma 
model in two dimensions (see e.g.~\cite{Perelomov}).\footnote{
In fact, $\Phi$ in (\ref{4.29}) takes values in the Grassmannian Gr($r,n$),
and Gr$(1,n)=\C P^{n-1}$.} 
For this reason, one can overcome the previously mentioned difficulty with 
constructing an action (or energy from the viewpoint of 2+1 dimensions) for
static configurations.
Moreover, on solutions obeying the BPS conditions~(\ref{4.30}) the
topological charge
\begin{equation}\label{4.35}
{\cal Q}\=2\pi\th \int\!\!\diff\h^1\diff\h^2\;
\textrm{Tr}\;\tr\ \Phi\,\bigl\{D_+\Phi\,, D_-\Phi \bigr\}
\end{equation}
is proportional to the action (BPS bound)
\begin{equation}\label{4.36}
S\=2\pi\th \int\!\!\diff\h^1\diff\h^2\;
\textrm{Tr}\;\tr\,\bigl[D_+\Phi\,, D_-\Phi \bigr]
\end{equation}
and is finite for algebraic functions $T_e$ and $T_o$.
Here, the standard superderivatives $D_\pm$ are defined as
\begin{equation}\label{4.34}
D_+\=\frac{\pa}{\pa\h} + \im\h\,\pa_z \und 
D_-\=\frac{\pa}{\pa\bar\h} + \im\bar\h\,\pa_{\bar z}\ .
\end{equation}
 
\smallskip

\noindent
{\bf One-soliton configuration. }
For one moving soliton, from (\ref{4.26}) and (\ref{4.27}) we obtain
\begin{equation}\label{4.37}
\psi_1\=\unity\ +\ \frac{\mu -\bar\mu}{\zeta -\mu}\,P 
\qquad\textrm{with}\qquad P\=T\,(T^\+ T)^{-1}T^\+
\end{equation}
and
\begin{equation}\label{4.38}
\Phi \= \unity\ -\ \rho\,P \qquad\textrm{with}\qquad 
\rho \= 1 - \frac{\mu }{\bar\mu}\ .
\end{equation}
Now our $n{\times}r$ matrix $T$ must satisfy 
(putting $Z^i=0$ and $Z=c\otimes\unity_{r\times r}$)
\begin{equation}\label{4.39}
[c\,,\,T]\=0 \und \frac{\pa }{\pa\bar\h^i}\,T\=0 
\qquad\textrm{with}\quad \h^i=\h_i^1+{\bar\m}\,\h_i^2\ ,
\end{equation}
where $c$ is the moving-frame annihilation operator given 
by~(\ref{4.15}) for $k{=}1$. 

Recall that the operators $c$ and $c^\+$ and therefore the matrix $T$ and the 
projector $P$ can be expressed in terms of the corresponding static objects 
by a unitary squeezing transformation (see e.g.~(\ref{4.10}) and~(\ref{4.15})).
For simplicity we again consider the case $\Ncal{=}1$ and a nonabelian 
projector with $r{=}1$.
Then (\ref{4.39}) tells us that $T$ is a holomorphic function of $c$ and $\h$,
i.e.
\begin{equation}\label{4.40}
T\= T_e(c)\ +\ \h\,T_o(c) \= \Biggl(\begin{smallmatrix} 
T_e^1(c)\ +\ \h\,T_o^1(c) \\[-4pt] \vdots \\[2pt] 
T_e^n(c)\ +\ \h\,T_o^n(c) \end{smallmatrix}\Biggr)
\end{equation}
with polynomials $T_e^a$ and $T_o^a$ of order~$q$, say,
analogously to the static case~(\ref{4.31}).
Note that, for $T_o^a$ to be Grassmann-odd and nonzero, 
some extraneous Grassmann parameter must appear.
Similarly, abelian projectors for a moving one-soliton obtain by subjecting
(\ref{4.31a}) to a squeezing transformation.

For ${\cal N}{=}1$ the moving frame was defined in (\ref{4.10}) 
(dropping the index $k$) via
\begin{equation}\label{4.42}
w\=x\ +\ \sfrac12(\bar\m{-}\bar\m^{-1})y\ +\ \sfrac12(\bar\m{+}\bar\m^{-1})t
\und \h\=\h^1 + \bar\m\h^2 \qquad\textrm{hence}\quad \pa_t\h =0\ .
\end{equation}
Consider the moving frame with the coordinates $(w,\bar w,s;\h,\bar\h)$ with
the choice $s=t$ and the related change of the derivatives 
(see~\cite{LP1, ChuLe})
\begin{subequations}\label{4.43}
\begin{eqnarray} 
\pa_x &=& \pa_w\ +\ \pa_{\bar w}\ ,\\
\pa_y &=& \sfrac12(\bar\m{-}\bar\m^{-1})\,\pa_w\ +\ 
          \sfrac12(\m{-}\m^{-1})\,\pa_{\bar w}\ ,\\
\pa_t &=& \sfrac12(\bar\m{+}\bar\m^{-1})\,\pa_w\ +\
          \sfrac12(\m{+}\m^{-1})\,\pa_{\bar w}\ +\ \pa_{s}\ ,\\
\pa_{\h^1} &=& \pa_{\h}\ +\ \pa_{\bar\h}\ ,\\
\pa_{\h^2} &=& \bar\m\,\pa_{\h}\ +\ \m\,\pa_{\bar\h}\ .
\end{eqnarray}
\end{subequations} 
In the moving frame our solution (\ref{4.38}) is static, 
i.e.~$\pa_{s}\Phi =0$, and the projector $P$ has the same form as 
in the static case. The only difference is the coefficient $\rho$ instead
of $2$ in~(\ref{4.38}). Therefore, by computing the action~(\ref{4.36}) 
in $(w,\bar w;\h^1,\h^2)$ coordinates, we obtain for algebraic functions~$T$
in~(\ref{4.40}) a finite answer, which differs from the static one by a 
kinematical prefactor depending on~$\m$ (cf.~\cite{LP1} for the bosonic case).

\smallskip

\noindent
{\bf Large-time asymptotics. }
Note that in the distinguished $(z, \bar z, t)$ coordinate frame 
(\ref{4.42}) implies that at large times $w\to \kappa\,t$ with 
$\kappa=\sfrac12({\bar\m}{+}{\bar\m}^{-1})$. As a consequence,
the $t^q$ term in each polynomial in 
(\ref{4.40}) will dominate, i.e.
\begin{equation}\label{4.44}
T \quad\to\quad t^q\ \Biggl(\begin{smallmatrix}
a_1\ +\ \h\,b_1 \\[-4pt] \vdots \\[2pt]
a_n\ +\ \h\,b_n \end{smallmatrix}\Biggr)
\ =:\ t^q\ \Gamma\ ,
\end{equation}
where $\Gamma$ is a fixed vector in $\C^n$.
It is easy to see that in the distinguished frame the large-time limit of 
$\Phi$ given by (\ref{4.38}) is
\begin{equation}\label{4.45}
\lim\limits_{t\to\pm\infty}\Phi \= \unity\ -\ \r\,\Pi 
\qquad\textrm{with}\qquad \Pi \= \Gamma\,(\Gamma^\+\Gamma)^{-1}\Gamma^\+
\end{equation}
being the projector on the constant vector $\Gamma$. 

Consider now the $m$-soliton configuration (\ref{4.27}). By induction of the 
above argument one easily arrives at the $m$-soliton generalization of 
(\ref{4.45}). Namely, in the frame moving with the $\ell$th lump we have
\begin{equation}\label{4.47}
\lim_{t\to\pm\infty}\Phi_m\=
(\unity-\r_1\Pi_1)\ldots(\unity-\r_{\ell-1}\Pi_{\ell-1})
(\unity-\r_\ell P_\ell)(\unity-\r_{\ell+1}\Pi_{\ell+1})\ldots(\unity-\r_m\Pi_m)
\ ,
\end{equation}
where the $\Pi_m$ are constant projectors. 
This large-time factorization of multi-soliton solutions provides
a proof of the no-scattering property because the asymptotic configurations
are identical for large negative and large positive times.

\vspace{5mm} 
 
\section{Conclusions} 
 
\noindent
In this paper we introduced a generalization of the modified integrable 
U($n$) chiral model with ${2\Ncal}{\le}\,8$ supersymmetries in 2+1 dimensions 
and considered a Moyal deformation of this model. 
It was shown that this $\Ncal$-extended chiral model is equivalent to 
a gauge-fixed BPS subsector of an $\Ncal$-extended super Yang-Mills model  
in 2+1~dimensions originating from twistor string theory. The dressing method 
was applied to generate a wide class of multi-soliton configurations, 
which are time-dependent finite-energy solutions to the equations of motion.
Compared to the $\Ncal{=}0$ model, the supersymmetric extension was seen
to promote the configurations' building blocks to holomorphic functions
of suitable Grassmann coordinates. By considering the large-time 
asymptotic factorization into a product of single soliton solutions we have 
shown that no scattering occurs within the dressing ansatz chosen here.

The considered model does not stand alone but is motivated by twistor string
theory~\cite{Wit} with a target space reduced to the mini-supertwistor 
space~\cite{group9, PSW, sigma8}. In this context, the obtained multi-soliton
solutions are to be regarded as D(0$\mid$2$\Ncal$)-branes moving inside 
D(2$\mid$2$\Ncal$)-branes~\cite{LeSa}. Here 2$\Ncal$ appears due to 
fermionic worldvolume directions of our branes in the superspace 
description~\cite{LeSa}. Switching on a constant $B$-field simply deforms 
the sigma model and D-brane worldvolumes noncommutatively, thereby admitting 
also regular supersymmetric noncommutative abelian solutions.

Restricting to static configurations, the models can be specialized to
Grassmannian supersymmetric sigma models, where the superfield~$\Phi$ takes
values in Gr($r,n$), and the field equations are invariant under 
2$\Ncal$~supersymmetry transformations with $0\le\Ncal\le 4$. 
This differs from the results for standard 2D sigma models~\cite{Zumino, AGF} 
where the target spaces have to be K\"ahler or hyper-K\"ahler for admitting 
two or four supersymmetries, respectively. This difference will be discussed 
in more details elsewhere.

We derived the supersymmetric chiral model in 2+1 dimensions through 
dimensional reduction and gauge fixing of the $\Ncal$-extended supersymmetric 
SDYM equations in 2+2~dimensions. Recall that for the purely bosonic case most 
(if not all) integrable equations in three and fewer dimensions can be 
obtained from the SDYM equations (or their hierarchy~\cite{MW}) by suitable 
dimensional reductions (see e.g.~\cite{Ward}--\cite{Ivanova:1992tk} and 
references therein). Moreover, this Ward conjecture~\cite{Ward} was extended 
to the noncommutative case (see e.g.~\cite{HamToda, Dimakis:2000tm}).
It will be interesting to consider similar reductions of the $\Ncal$-extended
supersymmetric SDYM equations (and their hierarchy~\cite{Wolf:2004hp}) to 
supersymmetric integrable equations in three and two dimensions generalizing 
earlier results~\cite{Gates}.

\bigskip

\section*{Acknowledgements}

\noindent
We acknowledge fruitful discussions with C.~Gutschwager.
This work was supported in part by the Deutsche Forschungsgemeinschaft~(DFG). 

\bigskip

\end{document}